\documentclass[12pt,epsf]{article}
\usepackage{graphicx}
\usepackage{epsfig}
\begin{document}

\def\a{\alpha}
\def\b{\beta}
\def\c{\varepsilon}
\def\d{\delta}
\def\e{\epsilon}
\def\f{\phi}
\def\g{\gamma}
\def\h{\theta}
\def\k{\kappa}
\def\l{\lambda}
\def\m{\mu}
\def\n{\nu}
\def\p{\psi}
\def\q{\partial}
\def\r{\rho}
\def\s{\sigma}
\def\t{\tau}
\def\u{\upsilon}
\def\v{\varphi}
\def\w{\omega}
\def\x{\xi}
\def\y{\eta}
\def\z{\zeta}
\def\D{\Delta}
\def\G{\Gamma}
\def\H{\Theta}
\def\L{\Lambda}
\def\F{\Phi}
\def\P{\Psi}
\def\S{\Sigma}

\def\o{\over}
\def\beq{\begin{eqnarray}}
\def\eeq{\end{eqnarray}}
\newcommand{\gsim}{ \mathop{}_{\textstyle \sim}^{\textstyle >} }
\newcommand{\lsim}{ \mathop{}_{\textstyle \sim}^{\textstyle <} }
\newcommand{\vev}[1]{ \left\langle {#1} \right\rangle }
\newcommand{\bra}[1]{ \langle {#1} | }
\newcommand{\ket}[1]{ | {#1} \rangle }
\newcommand{\EV}{ {\rm eV} }
\newcommand{\KEV}{ {\rm keV} }
\newcommand{\MEV}{ {\rm MeV} }
\newcommand{\GEV}{ {\rm GeV} }
\newcommand{\TEV}{ {\rm TeV} }
\def\diag{\mathop{\rm diag}\nolimits}
\def\Spin{\mathop{\rm Spin}}
\def\SO{\mathop{\rm SO}}
\def\O{\mathop{\rm O}}
\def\SU{\mathop{\rm SU}}
\def\U{\mathop{\rm U}}
\def\Sp{\mathop{\rm Sp}}
\def\SL{\mathop{\rm SL}}
\def\tr{\mathop{\rm tr}}

\def\IJMP{Int.~J.~Mod.~Phys. }
\def\MPL{Mod.~Phys.~Lett. }
\def\NP{Nucl.~Phys. }
\def\PL{Phys.~Lett. }
\def\PR{Phys.~Rev. }
\def\PRL{Phys.~Rev.~Lett. }
\def\PTP{Prog.~Theor.~Phys. }
\def\ZP{Z.~Phys. }

\newcommand{\bear}{\begin{array}}  
\newcommand {\eear}{\end{array}}
\newcommand{\la}{\left\langle}  
\newcommand{\ra}{\right\rangle}
\newcommand{\non}{\nonumber}  
\newcommand{\ds}{\displaystyle}
\newcommand{\red}{\textcolor{red}}
\def\ubl{U(1)$_{\rm B-L}$}
\def\REF#1{(\ref{#1})}
\def\lrf#1#2{ \left(\frac{#1}{#2}\right)}
\def\lrfp#1#2#3{ \left(\frac{#1}{#2} \right)^{#3}}
\def\OG#1{ {\cal O}(#1){\rm\,GeV}}


\baselineskip 0.7cm

\begin{titlepage}

\begin{flushright}
UT-09-22\\
IPMU 09-0124
\end{flushright}

\vskip 1.35cm
\begin{center}
{\large \bf
Relaxing a constraint on the number of messengers in a low-scale gauge mediation
}
\vskip 1.2cm
Ryosuke Sato$^{1,2}$, T. T. Yanagida$^{2,1}$
and Kazuya Yonekura$^{1,2}$ 
\vskip 0.4cm

{\it $^1$  Department of Physics, University of Tokyo,\\
   Tokyo 113-0033, Japan\\
$^2$ Institute for the Physics and Mathematics of the Universe (IPMU), 
University of Tokyo,\\ Chiba 277-8568, Japan}

\vskip 1.5cm

\abstract{ 
We propose a mechanism for relaxing a constraint on the number of messengers in  low-scale gauge mediation models. 
The Landau pole problem for the standard-model gauge coupling constants in the low-scale gauge mediation can be circumvented by using our mechanism.
An essential ingredient is a large positive anomalous dimension of messenger fields given by a large Yukawa coupling in a conformal field theory at high energies.
The positive anomalous dimension reduces the contribution of the messengers to the beta function
of the standard-model gauge couplings. }
\end{center}
\end{titlepage}

\setcounter{page}{2}

\section{Introduction}
\label{sec:1}

The low-energy-scale gauge mediation with the gravitino mass $m_{3/2} < O(10)$ eV is very attractive, since it does not suffer from any cosmological gravitino problem \cite{Viel:2005qj}. In such a low-scale gauge mediation, the messengers have their masses of the order $10^2 - 10^3$ TeV.
If the number of messengers, $N_{\rm mess}$, is large, the gauge coupling constants of the standard model (SM) easily blow up below the GUT scale, 
i.e. the gauge coupling constants hit  Landau poles below the GUT scale. 
The requirement of the perturbative unification of the SM gauge coupling constants, thus, leads to a constraint on the number of  messengers. 
It is known \cite{Jones:2008ib} that $N_{\rm mess} < 5$ for the messengers being ${\bf 5} + {\bf 5}^*$ of $SU(5)_{\rm GUT}$ if the masses of messengers
are smaller than about $10^3~\TEV$.

The above constraint becomes more severe if one considers  strongly interacting  messengers in
direct \cite{Giudice:1998bp} 
or semi-direct~\cite{Izawa:1997hu,Seiberg:2008qj} gauge mediation models, for instance.  
This is because the messengers receive most likely negative anomalous dimensions from the hidden strong gauge interactions and hence the SM gauge couplings run faster (see Section \ref{sec:2}).

In this paper we point out that it is not always the case if the theory is embedded into a conformal field theory at high energies. We show several examples where hidden sector interactions induce even positive large anomalous dimensions for messengers. In those example models one may have $N_{\rm mess} \ge 5$ without ruining the perturbative unification. 
A crucial ingredient is an introduction of a large Yukawa coupling of the messengers to some other hidden sector fields.

\section{Relaxing the constraint on $N_{\rm mess}$}
\label{sec:2}
In this section we describe our mechanism for relaxing the constraint on the number of messengers. 
In supersymmetric (SUSY) gauge theories, the $\b$ function of a gauge coupling is exactly given by the 
Novikov-Shifman-Vainshtein-Zakharov (NSVZ) $\b$ function~\cite{Novikov:1983uc}, 
\beq
\b(g)=\m \frac{\q}{\q \m} g^2=-\frac{g^4}{8\pi^2}\frac{3t(A)-\sum_{i} (1-\g_i)t(i)}{1-t(A)g^2/8\pi^2},\label{eq:NSVZ}
\eeq
where $t(A)$ and $t(i)$ are the Dynkin indices for the adjoint representation and the representation of matter fields $i$, 
$\g_i$ an anomalous dimension of matter $i$, and $\mu$ a renormalization scale.
Let us consider the $\b$ functions of the standard-model (SM) gauge couplings.
From the $\b$ function Eq.~(\ref{eq:NSVZ}), we can see that the effective messenger number contributing to the SM $\b$ functions is given by
\beq
N^{\rm eff}_{\rm mess} \equiv \sum_{i\in {\rm mess}} (1-\g_{i})t_{\rm GUT}(i), \label{eq:effmess}
\eeq
where the sum is taken over messenger fields, and $t_{\rm GUT}(i)$ are the Dynkin indices for 
the GUT gauge group~$SU(5)_{\rm GUT}$. Here, we have assumed that the messengers form a complete representation of the $SU(5)_{\rm GUT}$. 

It should be noted here that the $\b$ function and the gauge coupling constant, in fact, depend on renormalization schemes. 
The coupling $g_{\rm NSVZ}$ which satisfies Eq.~(\ref{eq:NSVZ}) is not the same as the one defined in 
a more conventional scheme, such as the dimensional reduction with minimal subtraction (DRED), $g_{\rm DRED}$.  
They are related (up to four loops) by the equations presented in Ref.~\cite{Jack:1998uj}. The relation between $g_{\rm NSVZ}$
and $g_{\rm DRED}$ may be given by~\cite{Jack:1998uj} 
\beq
\frac{g^2_{\rm DRED}}{16\pi^2}=\frac{g^2_{\rm NSVZ}}{16\pi^2}+C_{\rm loop}\left(\frac{g^2_{\rm NSVZ}}{16\pi^2}\right)^2,
\eeq
where $C_{\rm loop}$ is some loop-suppressed quantity which depends on couplings of the theory (including $g$ itself).
Thus, as long as the SM gauge couplings, $g^{\rm SM}$, are small enough, we may neglect the difference between $g_{\rm NSVZ}^{\rm SM}$ 
and $g_{\rm DRED}^{\rm SM}$, even if the messengers are strongly coupled by hidden interactions.
Thus, the description in terms of Eq.~(\ref{eq:NSVZ}) is well applicable for our purpose, since we examine a constraint on the number of the messengers
for maintaining the perturbative unification of the SM gauge coupling constants (i.e. $g_{\rm SM}^2/16\pi^2\ll1$).

Now let us suppose that the messengers are charged under some hidden gauge group (with the gauge coupling $g$) and have a Yukawa interaction
(with the Yukawa coupling $\l$) with some hidden matters. Then, the anomalous dimension of the messengers is given by, at the one-loop level, 
\beq
\g \sim -\frac{g^2}{16\pi^2}+\frac{|\l|^2}{16\pi^2}. \label{eq:roughanom}
\eeq
(Here, we have neglected the contributions from the SM gauge interactions.)
Then, from Eqs.~(\ref{eq:effmess}) and (\ref{eq:roughanom}), we see that the hidden gauge interaction increases the effective messenger 
number, $N^{\rm eff}_{\rm mess}$,
while the hidden Yukawa interaction decreases it.

In direct \cite{Giudice:1998bp} or semi-direct~\cite{Izawa:1997hu,Seiberg:2008qj} gauge mediation models and 
also in composite messenger models (e.g.~\cite{Hamaguchi:2007rb}), the messenger fields are supposed to be charged under hidden  
gauge groups with strong couplings. 
Thus, the messenger fields have negative
anomalous dimensions and the effective messenger number $N_{\rm mess}^{\rm eff}$ increases. 
Thus, the Landau pole problem discussed in the Introduction becomes more severe (for a more quantitative discussion, see \ref{app:B}).
However, if we introduce large Yukawa interactions in the messenger sector, the anomalous dimensions of the messengers can become positive and 
we can decrease the effective messenger number, $N^{\rm eff}_{\rm mess}$. For this mechanism to be efficient, it is desirable that the hidden gauge theory of the messengers is embedded into
a conformal field theory, because the messengers can have  large positive anomalous dimensions over a wide range of energy scales.
(Otherwise, the large Yukawa coupling hits its own Landau pole below the GUT scale.)

In the rest of this section we give example models which realize the above mechanism. We will see that the models have direct applications to low-scale gauge mediation models in the next section.

The models are based on  an $SU(N_C)$ hidden gauge group.
We first introduce $N_Q$ pairs of messenger quarks and antiquarks, $Q^i_\alpha$ and ${\tilde Q}^\alpha_i$, with $i=1,\cdots, N_Q$ and $\alpha=1,\cdots,N_C$.
The messengers $Q^i_\alpha$ and ${\tilde Q}^\alpha_i$ transform as fundamental and antifundamental representations of $SU(N_C)$, respectively.
We restrict our discussion to the case of $N_Q=5$, for simplicity, and assume that the quarks $Q^i_\alpha$ and antiquarks ${\tilde Q}^\alpha_i$ transform as ${\bf 5}^*$ and ${\bf 5}$ of $SU(5)_{\rm GUT}$, respectively. We introduce a mass term $m_Q Q^i_\alpha{\tilde Q}^\alpha_i$ for the messengers $Q^i_\alpha$ and ${\tilde Q}^\alpha_i$. Notice that $N_C$ is identified with the number of the messengers, $N_{\rm mess}$. The generalization to other gauge theories such as $SP(N_C)$ or $SO(N_C)$ is straightforward and hence we do not discuss it in this paper.

Now we introduce $N_P$ pairs of quarks and antiquarks, $P^{p}_\alpha$ and ${\tilde P}^{\alpha}_p$ with $p=1,\cdots,N_P$, and an adjoint quark chiral multiplet, $A^\alpha_\beta$ with $\alpha,~ \beta=1,\cdots,N_C$, to embed the theory into a superconformal field theory~\cite{Seiberg:1994pq} for giving the messengers  positive anomalous dimensions. 
We introduce their mass terms,
\begin{equation}
W_{\rm mass}= m_{P} P^{p}_\alpha {\tilde P}^{\alpha}_p +
m_{A} A^\alpha_\beta A_\alpha^\beta.
\end{equation}
We assume, $m_{P}, m_{A} > m_{Q},$
for the additional quarks, $P^{p}_\alpha$ and ${\tilde P}^{\alpha}_p$, and the adjoint quark $A^\alpha_\beta$ to decouple from the strong dynamics below the messenger mass scale.
We also introduce a Yukawa coupling,
\begin{equation}
W_{\rm Yukawa}= \sqrt{2}\l Q^{i}_\alpha A^\alpha_\beta {\tilde Q}^{\beta}_i .
\end{equation}
The introduction of the Yukawa coupling is important for our mechanism to work, as explained above.

\begin{table}[Ht]
\begin{center}
\begin{tabular}{|c|c|c|c|c|c|}
\hline 
&$N_P=2$&$N_P=3$&$N_P=4$&$N_P=5$&$N_P=6$ \\ \hline
$N_C=5$&0.303~(3.48)&0.156~(4.22)&0.062~(4.69)&$\times$&$\times$  \\ \hline
$N_C=6$&0.452~(3.29)&0.300~(4.20)&0.191~(4.85)&0.110~(5.34)&0.048~(5.71) \\ \hline
$N_C=7$&$\times$&0.411~(4.13)&0.301~(4.90)&0.214~(5.50)&0.144~(5.99) \\ \hline
$N_C=8$&$\times$&0.494~(4.05)&0.388~(4.89)&0.302~(5.58)&0.230~(6.16) \\ \hline
$N_C=9$&$\times$&$\times$&0.458~(4.88)&0.375~(5.63)&0.303~(6.27) \\ \hline
\end{tabular}
\caption{The anomalous dimensions $\g_Q$ of messenger fields $Q,~\tilde{Q}$ and the effective messenger numbers $N^{\rm eff}_{\rm mess}$ (in parentheses) at the conformal fixed points. Models marked with $\times$ do not have a desirable fixed point.}
\label{table:1} 
\end{center}
\end{table}

We find that the theory has an infrared conformal fixed point for a given appropriate value of $N_C$ and that of $N_P$. We show, in \ref{app:A}, the detailed determination of the infrared fixed points and of the anomalous dimensions of the messenger fields. We give the obtained anomalous dimensions $\g_Q$ of the messenger fields $Q$ and $\tilde{Q}$ and the effective messenger numbers 
$N^{\rm eff}_{\rm mess}=(1-\g_Q)N_C$ for various sets of $(N_C,~N_P)$ in Table~\ref{table:1}.
We see that  the models have the effective messenger numbers $N^{\rm eff}_{\rm mess}<5$ for many sets of  $(N_C,~N_P)$, 
even if the actual messenger number $N_{\rm mess}=N_C\geq 5$.

\section{Applications to low-scale gauge mediation models}
\label{sec:3}

In this section we discuss applications of our mechanism to various gauge mediation models. The applications have three categories, (I) application to a direct  gauge mediation, (II) that to a semi-direct gauge mediation and (III) that to a composite messenger model. We consider a representative model for each category to illustrate our mechanism discussed in Section~\ref{sec:2}.
\subsection{Direct gauge mediation}

Let us consider a direct gauge mediation model~\cite{Kitano:2006xg,Csaki:2006wi,Abel:2007jx,Haba:2007rj,Essig:2008kz}
in which a subgroup of the flavor symmetry of the Intriligator-Seiberg-Shih (ISS) model~\cite{Intriligator:2006dd} is gauged by $SU(5)_{\rm GUT}$.
The model is based on an $SU(N_C)$ gauge theory with $N_F$ pairs of
quarks $Q^I_\alpha$ and antiquarks ${\tilde Q}^\alpha_I$. Here, $I$ and $\alpha$ run from $I=1$ to $I=N_F$ 
and from $\alpha=1$ to $\alpha=N_C$, respectively. We assume, for simplicity, that they have a common mass
\begin{equation}
W=mQ^I_\alpha {\tilde Q}^\alpha_I.
\end{equation}
We have a global flavor symmetry $SU(N_F)_F$.

If the numbers of color and flavor satisfy the relation $N_C+1 \leq N_F<\frac{3}{2}N_C$,
this theory has a weakly coupled dual magnetic description at low energies. 
The dual magnetic theory is described in terms of mesons $\F^I_J$ and dual quarks $\v_I^{a},~\tilde{\v}^I_{a}$. 
Here, $a=1,\cdots, \tilde{N}_C$ is the index of a dual gauge group $SU(\tilde{N}_C=N_F-N_C)_{\rm mag}$. The superpotential of this theory is given by
\beq
W=h \v_I^{a}\F^I_J\tilde{\v}^J_{a}-h\mu^2 \F^I_I.\label{eq:magsuper}
\eeq
Without a loss of generality, the Yukawa coupling constant $h$ and the dimension one parameter $\m$ can be taken to be real and positive.

At the tree level, the equation of motion of $\F$ gives the $F$-term of $\F$,
\beq
-(F^\dagger_\F)_I^J=h \v_I^{a}\tilde{\v}^J_{a}-h\mu^2 \d^J_I.
\eeq
The right hand side of this equation cannot be zero, since the rank of the matrix $\v_I^{a}\tilde{\v}^J_{a}$ is no greater than $N_F-N_C$ and 
the unit matrix $\d^J_I$ has rank $N_F~(>N_F-N_C)$. Thus some components of $(F^\dagger_\F)_I^J$ are nonzero and SUSY is broken.
If nonperturbative effects of $SU(\tilde{N}_C)_{\rm mag}$ are taken into account, however, SUSY is dynamically restored~\cite{Intriligator:2006dd}. 
So the SUSY-breaking vacua are metastable. We consider a SUSY-breaking local minimum in the following discussion.

Around the SUSY-breaking local minimum of the potential, the fields $\v,\tilde{\v}$ and $\F$ can be expanded like
\beq
\v^a_I=\left(
\bear{cc}
\mu \d^a_p+\d\chi^a_p &
\d\r^a_i
\eear
\right),~~~
\tilde{\v}_a^I=\left(
\bear{c}
\mu \d^p_a+\d\tilde{\chi}_a^p \\
\d\tilde{\r}_a^i
\eear
\right),~~~
\F^I_J=\left(
\bear{cc}
\d Y^p_q &\d \tilde{Z}^p_j \\
\d Z_q^i&\d\F^i_j
\eear
\right),
\eeq
where $p=I$ for $1\leq I \leq N_F-N_C$ and $i=I$ for $N_F-N_C+1\leq I \leq N_F$. These vacuum expectation values (vevs) 
break the global flavor symmetry $SU(N_F)_F$
down to $SU(N_F-N_C)_F \times SU(N_C)_F$. To make this model a direct gauge mediation model, we embed the $SU(5)_{\rm GUT}$ gauge group into 
a subgroup of $SU(N_C)_F$ or $SU(N_F-N_C)_F$. Let us consider the theory above the mass scale $\m$ for each case~\cite{Kitano:2006xg}. 

\begin{enumerate} 
\item If $SU(5)_{\rm GUT} \subset SU(N_C)_F$, in the magnetic theory, fields charged under $SU(5)_{\rm GUT}$ are (a part of) $\d \r,~\d\tilde{\r},~\d Z,~\d\tilde{Z}$ in the (anti)fundamental representation of $SU(N_C)_F$ and $\d\F$ in the adjoint representation of $SU(N_C)_F$. 
Then, the contribution to the $\b$ function of the $SU(5)_{\rm GUT}$ gauge coupling is 
given by $N_{\rm mess}^{({\rm mag})}=2(N_F-N_C)+N_C=2N_F-N_C$~\footnote{$N_{\rm mess}$ is not equal to the one contributing to the gaugino and sfermion soft masses. In this paper we are defining $N_{\rm mess}$ only by the contribution to the gauge coupling $\b$ function.}. (Note that the adjoint representation of 
$SU(N_C)_F$ decomposes into an adjoint representation of $SU(5)_{\rm GUT}$, $N_C-5$ flavors of fundamental and antifundamental representations of 
$SU(5)_{\rm GUT}$, and some singlets.)
In the electric theory $N_{\rm mess}^{({\rm ele})}=N_C$. From the inequalities $N_C \geq 5$ and $N_C+1\leq N_F<\frac{3}{2}N_C$, we obtain
$N_{\rm mess}^{({\rm mag})}\geq7$ and $N_{\rm mess}^{({\rm ele})} \geq 5$.

\item If $SU(5)_{\rm GUT} \subset SU(N_F-N_C)_F$, in the magnetic theory, fields charged under $SU(5)_{\rm GUT}$ are (a part of) $\d\chi,~\d\tilde{\chi},~\d Z,~\d\tilde{Z}$ in the (anti)fundamental representation of $SU(N_F-N_C)_F$ and $\d Y$ in the adjoint representation of $SU(N_F-N_C)_F$. (This counting is applicable above the mass scale $\m$. Below $\m$, the SM gauge group is in the diagonal subgroup of $SU(N_F-N_C)_F\times SU(\tilde{N}_C)_{\rm mag}$.)
Then, the contribution to the $\b$ function of the $SU(5)_{\rm GUT}$ gauge coupling is 
given by $N_{\rm mess}^{({\rm mag})}=(N_F-N_C)+N_C+(N_F-N_C)=2N_F-N_C$. In the electric theory $N_{\rm mess}^{({\rm ele})}=N_C$. From the inequalities $N_F-N_C \geq 5$ and $N_C+1\leq N_F<\frac{3}{2}N_C$, we obtain
$N_{\rm mess}^{({\rm mag})}> 20$ and $N_{\rm mess}^{({\rm ele})} > 10$.
\end{enumerate}

In the case $SU(5)_{\rm GUT} \subset SU(N_F-N_C)_F$, $N_{\rm mess}$ is too large, so we concentrate on the case 
$SU(5)_{\rm GUT} \subset SU(N_C)_F$. In this model, SUSY breaking is mediated to the Minimal Supersymmetric Standard Model (MSSM) 
sector by the fields $\d \r,~\d\tilde{\r},~\d Z,~\d\tilde{Z}$.
The superpotential becomes
\beq
W&=&h \v_I^{a}\F^I_J\tilde{\v}^J_{a}-h\mu^2 \F^I_I \nonumber \\
&=&h\mu (\d \r^a_i \d Z_a^i+\d \tilde{\r}_a^i \d \tilde{Z}^a_i)+h\r_i^{a}\F^i_j\tilde{\r}^j_{a}-h\mu^2 \F^i_i+\cdots,
\eeq
where dots represent terms irrelevant for the gauge mediation.
$\F^i_i$ has a nonvanishing $F$-term, and then $\d\r$ and $\d\tilde{\r}$ have SUSY-breaking masses.
This is the type of gauge mediation studied in Ref.~\cite{Izawa:1997gs}.
$R$-symmetry breaking is rather nontrivial in this model and one has to consider some modification of the theory. See~\cite{Kitano:2006xg,Csaki:2006wi,Abel:2007jx,Haba:2007rj,Essig:2008kz} for details.

Even in the case $SU(5)_{\rm GUT} \subset SU(N_C)_F$, $N_{\rm mess}\geq 5$ in both the electric and magnetic theory. 
In fact, the messenger number $N_{\rm mess}$ is smaller in the electric theory than in the magnetic theory, which was considered as a solution 
to the Landau pole problem in Ref.~\cite{Abel:2008tx}. However, the analyses of Ref.~\cite{Jones:2008ib} suggest that 
the two-loop effects from the MSSM sector make it difficult to maintain 
the perturbative GUT unification for the low-scale gauge mediation. 
Furthermore, the messenger fields are charged under the strong hidden gauge group $SU(N_C)$ in the electric theory~\footnote{
In the magnetic theory, $SU(5)_{\rm GUT}$ charged fields have both the $SU(\tilde{N}_C)_{\rm mag}$ gauge interaction (if $\tilde{N}_C \geq 2$) and the Yukawa interaction in the superpotential (\ref{eq:magsuper}). Then it is nontrivial whether the total effect of these interactions decreases or increases the effective messenger number. 
}. Thus as explained in
Section~\ref{sec:2}, this model suffers from the severe Landau pole problem when the messenger mass scale is of order $10^{5}~\GEV$.
Notice that such a small mass $\sim 10^5~\GEV$ for the messenger
is required in the models~\cite{Csaki:2006wi,Abel:2007jx,Haba:2007rj,Essig:2008kz}, since the MSSM gaugino masses vanish at the leading order of the
SUSY-breaking scale~\footnote{There are some models in which the gaugino masses are generated at the leading order of the 
SUSY-breaking scale, by ``uplifting 
metastable vacua''~\cite{Giveon:2009yu}, but even in such models the Landau pole problem may be
sometimes problematic.} (see also Ref.~\cite{Komargodski:2009jf}).

We now consider a modification of the model to avoid the Landau pole problem.
In the electric theory, we add a chiral field $A^\a_\b$ which transforms in the adjoint representation of the $SU(N_C)$ gauge group.
We introduce new terms in the superpotential
\beq
W \supset \sqrt{2}\l Q^i_\a A^\a_\b \tilde{Q}_i^\b+m_A A^\a_\b A^\b_\a,
\eeq
where $i=N_F-N_C+1,\cdots,N_F$. This is in fact the model considered in Section~\ref{sec:2}, with the identification
$P^p_\a|_{{\rm Section~\ref{sec:2}}}=Q^p_\a~~(p=1,\cdots,N_F-N_C)$, $N_Q|_{{\rm Section~\ref{sec:2}}}=N_C$ and
$N_P|_{{\rm Section~\ref{sec:2}}}=N_F-N_C$. We take $N_C=5$ and 
$2\leq N_F-N_C \leq 4$ in the following discussion. 

The dynamics of the model is as follows. At high energies, we assume that the theory is near the conformal fixed point. Then, as discussed in
Section~\ref{sec:2}, the effective messenger number $N_{\rm mess}^{\rm eff}$ is smaller than 5 (see Table~\ref{table:1}). Below the mass scale $m_A$,
the adjoint field $A$ decouples from the dynamics, and the theory exits from the conformal fixed point and the confinement occurs. Then at the low energies 
the model can be described by the weakly coupled magnetic theory. SUSY is broken as in
the ISS model, and the direct gauge mediation works. 

However, the low energy theory is not completely the same as the original ISS model.
Integration of the adjoint field $A$ generates a superpotential 
\beq
W\supset -\frac{\l^2}{2m_A}\left[(Q_\a^i\tilde{Q}^\a_j) (Q_\b^j\tilde{Q}^\b_i)-\frac{1}{N_C}  (Q_\a^i\tilde{Q}^\a_i) (Q_\b^j\tilde{Q}^\b_j) \right] 
= -\frac{\l^2 \L^2}{2m_A}\left[\d\F^i_j\d\F^j_i-\frac{1}{N_C}\d\F^i_i\d\F^j_j \right],\nonumber \\ \label{eq:higerop}
\eeq
where $\L$ is the confinement scale of the electric theory defined by $Q_\a^i\tilde{Q}^\a_j=\L \F^i_j$. 
When $N_C=N_Q|_{\rm Section~\ref{sec:2}}=5$, this term gives mass $\l^2 \L^2/m_A$ to the traceless part of $\d\F$, i.e. the part which transforms in the adjoint
representation of $SU(N_C)_F=SU(5)_{\rm GUT}$. 
The traceless part of $\d\F$ does not take part in SUSY breaking and gauge mediation, but this field contributes to 
the $\b$ functions of the SM gauge coupling constants. 
So the ``messenger number'' contributing to the $\b$ function is $N_{\rm mess}=2(N_F-N_C)$ below the mass scale $\l^2 \L^2/m_A$.
Thus, if we take $N_F-N_C=2$, the messenger number $N_{\rm mess}$ is smaller than 5 for the energy scale below $\l^2 \L^2/m_A$
in the magnetic theory. On the other hand,  
above the scale $m_A$, the theory is electric and
$N_{\rm mess}^{\rm eff}$ is small because of the mechanism of Section~\ref{sec:2}. 
Furthermore, $\L$ is roughly related to $m_A$ by the equation $\L \sim m_A \exp(-8\pi^2/(3N_C-N_F)g_*^2)$, where $g_*$ is the gauge coupling
constant of $SU(N_C)$ at the fixed point. Then, if the fixed point is strongly coupled, which is the case in the present model, $m_A$ and $\l^2 \L^2/m_A$
are of the same order. Thus the dangerous energy scale between $\l^2 \L^2/m_A$ and $m_A$ is narrow and the perturbative unification of the SM gauge 
couplings is maintained.

\subsection{Semi-direct gauge mediation}
We consider a SUSY-breaking model based on an $SU(5)_{\rm hid}$ gauge symmetry. 
It is known~\cite{Affleck:1983vc} that the SUSY is broken when we introduce only two matter multiplets,
$V^\a$ and $X_{\a\b}$ in the representations ${\bf 5}^*$ and ${\bf 10}$ of $SU(5)_{\rm hid}$, respectively. 
We now introduce $N_F$ pairs of
fundamental quarks $Q^i_\alpha$ and antiquarks ${\tilde Q}^\alpha_i$. Here, $i=1,\cdots ,N_F$ and $\alpha=1,\cdots,5$ and
they belong to $({\bf N_F^*, 5})$ and $({\bf N_F, 5^*})$ representations of the 
$SU(N_F)_F \times SU(5)_{\rm hid}$, respectively.
We introduce a common bare mass term for the messengers, for simplicity,
\beq
W=m_Q Q^i_\alpha {\tilde Q}^\alpha_i.
\eeq

We gauge a subgroup of $SU(N_F)$ by the $SU(5)_{\rm GUT}$ gauge group~\cite{Ibe:2007wp}.
This is the setup for the semi-direct gauge mediation in the $SU(5)_{\rm hid}$ SUSY-breaking model~\footnote{
In semi-direct gauge mediation, a messenger number is not necessarily larger than or equal to 5.
For example, in the models of Ref.~\cite{Seiberg:2008qj}, the messenger number is minimally 2, so there is no Landau pole problem.
However, in semi-direct gauge mediation, sparticle masses (especially gaugino masses) are suppressed by hidden sector loops~\cite{Ibe:2007wp,Seiberg:2008qj},
so if one wants to build a model in which the gravitino is very light ($<{\cal O}(10)~\EV$), the hidden sector gauge theory should be strongly coupled.  
Because the gauge theory should be strongly coupled even when we add messenger fields, 
the hidden sector gauge group should be somehow large,
as in the above $SU(5)_{\rm hid}$ model. Then the model may suffer from the Landau pole problem. 
}.
The bifundamental messenger fields $Q$ and $\tilde{Q}$ link the $SU(5)_{\rm hid}$ hidden gauge sector and 
the MSSM sector, thus SUSY-breaking is mediated to the MSSM sector. In particular, 
when $N_F\geq6$, it can be shown that the theory has an infrared conformal fixed point
above the mass scale $m_Q$. Then, after the decoupling of the messengers $Q$ and $\tilde{Q}$, the theory exits from the conformal fixed point and the SUSY breaking occurs. This is a conformal gauge
mediation model proposed in Ref.~\cite{Ibe:2007wp}. However, we only impose $N_F \geq5$ in this paper.

In the above model, the messenger fields are charged under the strong $SU(5)_{\rm hid}$ gauge group. 
Thus, as discussed in Section~\ref{sec:2}, the effective messenger
number $N_{\rm mess}^{\rm eff}$ is larger than 5. Thus this theory suffers from the Landau pole problem.
To avoid the problem, we introduce a chiral field $A^\a_\b$ transforming in the adjoint representation of $SU(5)_{\rm hid}$. We introduce a superpotential,
\beq
W \supset \sqrt{2}\l Q^i_\a A^\a_\b \tilde{Q}_i^\b+m_A A^\a_\b A^\b_\a,
\eeq
for the mechanism explained in Section~\ref{sec:2} to work. In fact, this model is not the same as the example model described in Section~\ref{sec:2},
but we can study the infrared conformal fixed point of this theory by using the $a$-maximization technique explained in \ref{app:A}.
The result is listed in Table~\ref{table:2}. One can see that there is no Landau pole problem for $N_F=5,~6$ and $7$.

\begin{table}[Ht]
\begin{center}
\begin{tabular}{|c|c|c|c|c|c|}
\hline 
$N_F$&$\g_Q$&$\g_A$&$\g_V$&$\g_X$&$N_{\rm mess}^{\rm eff}$ \\ \hline
5&$0.264$&$-0.528$&$-0.656$&$-0.901$&3.68  \\ \hline
6&$0.160$&$-0.320$&$-0.527$&$-0.731$&$4.20$ \\ \hline
7&$0.063$&$-0.126$&$-0.308$&$-0.438$&$4.69$ \\ \hline
\end{tabular}
\caption{The anomalous dimensions of the fields of the model. $\g_Q$ is the anomalous dimension of $Q$ and $\tilde{Q}$. $\g_A,~\g_V$ and $\g_X$ are the anomalous dimensions of $A,~V$ and $X$ respectively. The fact that the anomalous dimensions of $Q$ and $\tilde{Q}$ are the same is not obvious because the model is chiral, but $a$-maximization shows that is the case. $N_{\rm mess}^{\rm eff}$ is defined by $N_{\rm mess}^{\rm eff}=5(1-\g_Q)$.}
\label{table:2} 
\end{center}
\end{table}

The dynamics of the model is as follows. We take $m_A>m_Q$. Then we assume that the theory is near the conformal fixed point above the mass $m_A$.
Below the threshold of $A$, the $SU(5)_{\rm hid}$ gauge coupling constant becomes larger, and it blows up (when $N_F=5$) or goes to another
fixed point discussed above (when $N_F \geq6$). In any case, after the decoupling of the messenger fields $Q$ and $\tilde{Q}$, 
SUSY is broken~\cite{Murayama:1995ng,Affleck:1983vc}.

However, the low energy theory after the decoupling of $A$ is not the same as the original semi-direct gauge mediation of Ref.~\cite{Ibe:2007wp}. 
As in Eq.~(\ref{eq:higerop}) of 
the previous subsection, the integration of $A$ generates a superpotential
\beq
W\supset -\frac{\l^2}{2m_A}\left[(Q_\a^i\tilde{Q}^\a_j) (Q_\b^j\tilde{Q}^\b_i)-\frac{1}{5}  (Q_\a^i\tilde{Q}^\a_i) (Q_\b^j\tilde{Q}^\b_j) \right] .\label{eq:explicR}
\eeq
The presence of this superpotential is very interesting, since this term explicitly breaks $R$ symmetry of the original semi-direct gauge mediation
model, which may be useful for a generation of gaugino masses~\cite{Ibe:2009bh}. 
To see that, we examine the original superpotential with the adjoint field $A$,
\beq
W=m_Q Q^i_\alpha {\tilde Q}^\alpha_i+\sqrt{2}\l Q^i_\a A^\a_\b \tilde{Q}_i^\b+m_A A^\a_\b A^\b_\a.
\eeq
Then the mass matrix of the quarks $Q,\tilde{Q}$ is given by $(m_Q\d^\a_\b+\sqrt{2}\l A^\a_\b)\d^i_j$.
Integrating the quarks, we obtain the following gauge kinetic term for the SM subgroup of $SU(5)_{\rm GUT}$:
\beq
&&-\int d^2\h \frac{1}{32\pi^2}\log\det[(m_Q\d^\a_\b+\sqrt{2}\l A^\a_\b)/\mu]{\cal W}^2_{\rm SM} \nonumber \\
&\simeq&\int d^2\h \frac{1}{32\pi^2}\left\{-\log(m_Q/\mu)^5+\frac{\l^2}{m_Q^2}A^\a_\b A^\b_\a+\cdots \right\}  {\cal W}^2_{\rm SM},
\eeq
where ${\cal W}_{\rm SM}$ is the SM gauge field strength normalized to have a kinetic term $\int d^2\h (1/4g_{\rm SM}^2){\cal W}^2_{\rm SM}$. 
Then we obtain the mass term for $A$,
\beq
m_A\left(1+\frac{\l^2{\cal W}^2_{\rm SM}}{32\pi^2m_Am_Q^2}\right) A^\a_\b A^\b_\a.
\eeq
Integrating out $A$, we finally obtain a holomorphic interaction term,
\beq
-\int d^2\h \frac{5}{4}\left(\frac{1}{16\pi^2}\right)^2\frac{\l^2}{m_Am_Q^2} {\cal W}^2_{\rm hid}{\cal W}^2_{\rm SM}.\label{eq:holointeraction}
\eeq
At tree level, this term generates the gaugino masses of order
\beq
M_{\tilde{g}} \sim \frac{5}{2}\left(\frac{g_{\rm hid}^2}{16\pi^2}\right)\left(\frac{g_{\rm SM}^2}{16\pi^2}\right)\frac{\l^2 D_{\rm hid}^2}{m_Am_Q^2},
\eeq
where $D_{\rm hid}$ is the $D$-term of $SU(5)_{\rm hid}$ normalized as ${\cal W}^2_{\rm hid}=g^2_{\rm hid}D^2_{\rm hid}\h^2+\cdots$, i.e. the vacuum 
energy is given by $V=D^2_{\rm hid}/2$.
Because the original $SU(5)_{\rm hid}$ model with $V^\a({\bf 5}^*)+X_{\a\b}({\bf 10})$  has no superpotential, it seems plausible that the $D$-term is responsible 
for the SUSY breaking, although the strong dynamics make it difficult to prove it. 
Then we have $D_{\rm hid} \neq 0$, and thus nonzero gaugino masses are obtained.
Note that the gaugino masses may be generated even without the explicit $R$ breaking term in Eq.~(\ref{eq:explicR}), since $R$ symmetry is 
believed to be spontaneously
broken in the $SU(5)_{\rm hid}$ model~\cite{Affleck:1983vc}. But without the explicit breaking, 
there are no holomorphic terms like Eq.~(\ref{eq:holointeraction}) and hence the mechanism  for the gaugino mass generation is not clear.

On the other hand, the term in Eq.~(\ref{eq:explicR}) generates SUSY preserving vacua
at $\vev{Q\tilde{Q}} \sim m_Am_Q/\l^2$, and hence the SUSY-breaking vacuum at the origin $\vev{Q\tilde{Q}}=0$ becomes metastable.
However, the SUSY-breaking vacuum decays only through vacuum tunneling, 
since the original semi-direct gauge mediation of Ref.~\cite{Ibe:2007wp} has no flat direction.
In the decoupling limit $m_A \to \infty$ with the low energy physics of Ref.~\cite{Ibe:2007wp} fixed~\footnote{
In the cases $N_F \geq 6$, the low energy physics, including the dynamical scale of the theory, is determined only by $m_Q$ in the limit $m_A \to \infty$.
On the other hand, if $N_F=5$, the dynamical scale is sensitive also to $m_A$, and the limit $m_A \to \infty$ must be accompanied with the limit 
$g_{\rm hid} \to 0$ which makes the theory off the conformal fixed point at high energies.}
, the SUSY-preserving vacua go to infinity, 
and the vacuum tunneling rate from the SUSY breaking vacuum to the SUSY preserving vacua should become negligible.
It is expected that the vacuum tunneling rate $\Gamma$ may be of the form,
\beq
\Gamma &\propto& e^{-S},  \\
S &\sim& c \left(\frac{m_A}{m_Q}\right)^a, \label{eq:classicalaction}
\eeq
where $c$ and $a$ are some dimensionless positive constants of ${\cal O}(1)$. We have assumed that the dynamical scale of the model is comparable to $m_Q$.
Because of the exponential factor,
$\Gamma$ is expected to become negligible quickly as we make $m_A$ large. 
If a so-called thin-wall approximation~\cite{Coleman:1977py} is applicable (which may be justified by the nonexistence of flat directions), we get
\beq
S \simeq \frac{27\pi^2}{2} \frac{S_1^4}{\e^3}, \label{eq:thin-wall-app}
\eeq
where $\e$ is the difference of vacuum energy between the metastable and true vacuum, $\e \sim m_Q^4$, 
and $S_1$ is given by
\beq
S_1 &\sim& \int_0^{\sqrt{m_Am_Q}} dQ \sqrt{2V(Q)} \nonumber \\
&\sim& m_Am_Q^2,
\eeq
where $V(Q)$ is the potential of $Q,\tilde{Q}$. This result gives $a=4$ in Eq.~(\ref{eq:classicalaction}).
If $\G \sim (10^5~\GEV)^4 \times e^{-S}$, we require $S \gsim 400$ for the lifetime of the vacuum to be sufficiently long, i.e. 
$\G \ll H_0^{4}$ where $H_0 \sim 10^{-42}~\GEV$ is the present value of the Hubble constant. 
Eq.~(\ref{eq:classicalaction}) leads to $S \gsim 400$ for $m_A/m_Q \gsim 5$ if $c \simeq 1$, 
but notice that the factor $27\pi^2/2 \simeq 133$ in Eq.~(\ref{eq:thin-wall-app}) may make $c$ much larger than 1, and a smaller ratio of $m_A/m_Q$ may 
be allowed.

\subsection{Composite messenger model}
We consider a strong $SU(5)_{\rm hid}$ gauge theory with 5 pairs of
fundamental quarks $Q^i_\alpha$ and antiquarks ${\tilde Q}^\alpha_i$. Here, $i$ and $\alpha$ run
from 1 to 5 and they belong to $({\bf 5^*, 5})$ and $({\bf 5, 5^*})$ representations of the 
$SU(5)_{\rm GUT}\times SU(5)_{\rm hid}$, respectively. Those 5 pairs of quarks play a role of messengers. 
We take a superpotential for the messengers,
\beq
W=h XQ^i_\alpha {\tilde Q}^\alpha_i, \label{eq:minimGM}
\eeq
where $X=M+F\h^2$ is a SUSY-breaking spurion field. This model is the so-called minimal gauge mediation \cite{Giudice:1998bp}, 
aside from the fact that the messengers are charged under
the strong gauge group $SU(5)_{\rm hid}$.

The reason why we introduce the $SU(5)_{\rm hid}$ gauge interaction is to confine the messenger quarks and antiquarks, $Q$ and $\tilde{Q}$, forming composite fields. One of the composite states can be a candidate for the dark matter of the Universe~\cite{Hamaguchi:2007rb}. In fact we have, at low energies, mesons 
\beq
M^i_j=Q^i_\a \tilde{Q}^\a_j,
\eeq
and baryons
\beq
B=\det Q,~~~\tilde{B}=\det\tilde{Q},
\eeq
with the constraint
\beq
\det M -B\tilde{B} =\L^{10},
\eeq
where $\L$ is the dynamical scale of $SU(5)_{\rm hid}$. The baryon $B$ and antibaryon $\tilde{B}$ are long-lived, since we have an approximate 
baryon number conservation~\cite{Hamaguchi:2007rb}. Then, they can be a candidate for the dark matter.

Above the energy scale $\L$, the messenger number is 5, but because the messengers are charged under the strong gauge group, the effective messenger
number $N_{\rm mess}^{\rm eff}$ is larger than 5 as explained in Section~\ref{sec:2}. Below the energy scale $\L$, the traceless part of $M^i_j$ transforms in the adjoint representation
of $SU(5)_{\rm GUT}$, so the messenger number (in the definition of this paper) is also 5. Thus this model suffers from the Landau pole problem.

For the mechanism of Section~\ref{sec:2} to work, we introduce additional $N_P$ pairs of quarks $P^p_\a$ and $\tilde{P}_p^\a$ in the representation $\bf{5}$
and $\bf{5}^*$ of $SU(5)_{\rm hid}$, respectively. Here $p$ is the flavor index, $p=1,\cdots,N_P$. We further introduce an adjoint field $A^\a_\b$ of 
$SU(5)_{\rm hid}$, and introduce a superpotential
\beq
W \supset \sqrt{2}\l Q^{i}_\alpha A^\alpha_\beta {\tilde Q}^{\beta}_i+m_{P} P^{p}_\alpha {\tilde P}^{\alpha}_p +
m_{A} A^\alpha_\beta A_\alpha^\beta,
\eeq 
for the additional fields, as in Section~\ref{sec:2}. Then the messenger model becomes the same as the model in Section~\ref{sec:2}.
The effective messenger number above the mass $m_P$ and $m_A$ is given in Table~\ref{table:1} with $N_C$ equal to 5.
We see that the Landau pole problem can be avoided by taking $m_P$ and $m_A$ appropriately small.

The dynamics of the model is as follows. We assume that $m_P$ and $m_A$ are of the same order, $m_P \sim m_A$, for simplicity, 
and the theory is near the conformal fixed point above the threshold of those fields. 
After the decoupling of $P,~\tilde{P}$ and $A$, the $SU(5)_{\rm hid}$ gauge coupling becomes strong
and the gauge theory confines the color degrees of freedom, making composite fields described above.

However, we have to take care of the following point. When the theory is on the conformal fixed point, the Yukawa coupling in Eq.~(\ref{eq:minimGM}) becomes smaller as we lower the renormalization scale. Suppose that the theory is
on the conformal fixed point from the energy scale $M_*$ down to $m_*~(\sim m_A \sim m_P)$. Then, neglecting all effects other than the fixed point dynamics,
the Yukawa coupling $h$ at the scale $m_*$ is
\beq
h |_{m_*} \equiv h_* \simeq \left(\frac{m_*}{M_*}\right)^{\g_Q} h |_{M_*}.
\eeq

We show that the requirement $m_{3/2}<16~\EV$~\cite{Viel:2005qj} leads to a constraint on the number of the additional quarks $P$ and $\tilde{P}$, $N_P$. 
For the messenger quarks not to be tachyonic, 
the SUSY-breaking scale $F$ must satisfy $h_*F<(h_*M)^2$. Then, the gaugino mass is constrained as~\footnote{The messenger fields are strongly coupled in the present model, but here we pretend as if they can be treated as weakly coupled fundamental quarks, for simplicity.
In fact, it is known that the gaugino masses are not so affected by strong interactions, due to the gaugino screening mechanism~\cite{ArkaniHamed:1998kj}.}
\beq
M_{\tilde{g}} \simeq n\frac{\a}{4\pi}\frac{h_*F}{h_*M} < n\frac{\a}{4\pi}\sqrt{h_*F},
\eeq
where $\a$ is the SM gauge coupling fine structure constant corresponding to the gaugino $\tilde{g}$, and $n$ a ``messenger number'' contributing to the gaugino masses (in the present model, $n=5$).
Then, the gravitino mass is constrained as
\beq
m_{3/2}=\frac{F}{\sqrt{3}M_{Pl}} > \frac{\left(4\pi \a^{-1} n^{-1} M_{\tilde{g}}\right)^2}{\sqrt{3}h_*M_{Pl}}=16~\EV \left(\frac{3.4\times 10^{-3}}{h_*}\right)
\left(\frac{\a^{-1}}{60}\frac{M_{\tilde g}}{100~\GEV}\right)^2,
\eeq
where $M_{Pl} \simeq 2.4 \times 10^{18}~\GEV$ is the reduced Planck mass. Note the dependence $1/h_*$ of this lower bound. 
Thus, to achieve the light gravitino mass, $h_*$ should not be too small,
and thus $\g_Q$ should not be too large. In particular, the model with $N_P=2$ may not be favored, although the effective messenger number 
$N_{\rm mess}^{\rm eff}$ is the smallest in this case.

\section*{Acknowledgement}
We would like to thank S. Shirai for important suggestions, and for very useful discussions and comments.
This work was supported by World Premier International Center
Initiative (WPI Program), MEXT, Japan.  The work of KY is supported in
part by JSPS Research Fellowships for Young Scientists.

\appendix
\setcounter{equation}{0}
\renewcommand{\theequation}{\Alph{section}.\arabic{equation}}
\renewcommand{\thesection}{Appendix~\Alph{section}}

\section{Effective messenger number in asymptotically free gauge theories} \label{app:B}
\setcounter{equation}{0}

In this appendix we study how strong gauge interactions make the Landau pole problem of the SM gauge couplings severe.
Suppose that messenger fields transform under the representation $r+\bar{r}$ of some strong gauge group $G$. At the one-loop level, the gauge coupling $g$ of the gauge group $G$ is given by 
\beq
\frac{8\pi^2}{g^2(\mu)}=\frac{8\pi^2}{g^2_0}+b\log(\m/M_0),
\eeq
where $b$ is the coefficient of the one-loop $\b$ function, and $g_0$ is the gauge coupling at the scale $M_0$.
We define $t \equiv \log(\mu/M_0)$ and $g^2(\mu)/8\pi^2 \equiv h_g(\mu)$ for simplicity. Then the above equation is rewritten as
\beq
h_g(t)=(h_{g0}^{-1}+bt)^{-1}.
\eeq
The anomalous dimension of the messenger fields is, at the one-loop level, given by
\beq
\g(t)=-2C_2(r)h_g(t)=-2C_2(r)(h_{g0}^{-1}+bt)^{-1},
\eeq
where $C_2(r)$ is the quardratic Casimir of the representation $r$.

We define the averaged  value of the anomalous dimension between $\mu=M_0$ and $\mu=M_1$ as
\beq
\tilde{\g} \equiv \frac{1}{t_1}\int^{t_1}_0 dt \g(t)=-\frac{2C_2(r)}{b} \frac{\log(1+bh_{g0}t_1)}{t_1}, 
\eeq 
where $t_1=\log(M_1/M_0)$. Then the averaged effective messenger number is 
\beq
\tilde{N}_{\rm mess}^{\rm eff}=(1-\tilde{\g})N_{\rm mess},
\eeq
where $N_{\rm mess}$ is ``the tree level value'' of the messenger number.

For example, consider the case that the hidden gauge group is $G=SU(5)_{\rm hid}$, the matter representation under $SU(5)_{\rm hid}$ is $r={\bf 5}$, 
the $\b$ function coefficient $b=3\times 5-5=10$, 
$M_0 =M_{\rm mess} \sim 10^6~\GEV$, and
$M_1=M_{\rm GUT} \sim 10^{16}~\GEV$. If we further assume that the coupling is very strong at $M_0$, e.g. $\g(\mu=M_0) \simeq -1$, 
we obtain
\beq
\tilde{\g} \simeq  -0.080,~~~~~\tilde{N}_{\rm mess}^{\rm eff} = (1-\tilde{\g})5 \simeq 5.40.
\eeq
The SM gauge couplings receive a contribution $\tilde{N}_{\rm mess}^{\rm eff}\log(M_{\rm GUT}/M_{\rm mess})$ from the messenger fields.
Defining $M'_{\rm eff}$ by the equation
\beq
\tilde{N}_{\rm mess}^{\rm eff}\log(M_{\rm GUT}/M_{\rm mess})=N_{\rm mess}\log(M_{\rm GUT}/M'_{\rm mess}),
\eeq
we obtain
\beq
\frac{M_{\rm mess}}{M'_{\rm mess}}=\left(\frac{M_{\rm GUT}}{M_{\rm mess}}\right)^{\frac{\tilde{N}_{\rm mess}^{\rm eff}}{N_{\rm mess}}-1} \sim 6.
\eeq
We see that the lower bound on the messenger mass scale becomes higher by this factor due to strong gauge interactions.

\setcounter{equation}{0}
\section{Details on conformal fixed point}\label{app:A}
In this appendix, we describe how to find a conformal fixed point and compute anomalous dimensions of matter fields, taking the model of Section \ref{sec:2} as an
example. We do not restrict the number of flavors of $Q$, $N_Q$, equal to $5$ in this appendix.

Let us first discuss the existence of the infrared fixed point in the model in Section~\ref{sec:2}.
For a time being we consider only the theory where the perturbative calculation for the $\b$ function is reliable.  
At the perturbative level, one can discuss the existence of a conformal fixed point by explicitly considering the renormalization group equations, as was first
done in Ref.~\cite{Banks:1981nn}.
The NSVZ $\b$ function of the gauge coupling $g$ and the $\b$ function of the Yukawa coupling $\l$ of the model in Section \ref{sec:2} are given by
\beq
\b_g&=&\m \frac{\q}{\q \m} g^2=-\frac{g^4}{8\pi^2}\frac{3N_C-(1-\g_A)N_C-(1-\g_Q)N_Q-(1-\g_P)N_P}{1-N_Cg^2/8\pi^2}, \\
\b_\l&=&\m \frac{\q}{\q \m} \l^2=(\g_A+2\g_Q)\l^2,
\eeq
where we have taken $\l$ to be real without a loss of generality, and $\g_A$, $\g_Q$ and $\g_P$ are the anomalous dimensions of $A$, $Q$, and $P$, respectively. At the one-loop level, they are given by~\footnote{See e.g. Section 5.5 of Ref.~\cite{Martin:1997ns}. Note that our convention for the anomalous dimension is larger than that used in Ref.~\cite{Martin:1997ns} by a factor of 2.}
\beq
\g_A \simeq \frac{N_Q\l^2-2N_Cg^2}{8\pi^2} ,~~~~~ \g_Q \simeq  \frac{N_C^2-1}{N_C}\frac{\l^2-g^2}{8\pi^2},~~~~~\g_P \simeq -\frac{N_C^2-1}{N_C}\frac{g^2}{8\pi^2}.
\eeq
By taking a large $N$ limit, $N_C,~N_Q,~N_P \gg 1$ with $N_Q/N_C \sim {\cal O}(1)$, $N_P/N_C \sim {\cal O}(1)$ and $n\equiv 3N_C-N_C-N_Q-N_P \sim {\cal O}(1)$, one can find a solution to the equations $\b_g=0,~\b_\l=0$ with the
couplings $g$ and $\l$ being very small (Banks-Zaks-like fixed point). The result is 
\beq
\frac{\l^2}{8\pi^2} \simeq \frac{4n}{(2N_C-N_Q)^2+N_P(2N_C+N_Q)},~~~~~\frac{g^2}{8\pi^2} \simeq \frac{N_Q+2N_C}{4N_C} \frac{\l^2}{8\pi^2},
\eeq 
and
\beq
\g_Q \simeq \frac{(2N_C-N_Q)n}{(2N_C-N_Q)^2+N_P(2N_C+N_Q)},~~\g_P \simeq -\frac{(2N_C+N_Q)n}{(2N_C-N_Q)^2+N_P(2N_C+N_Q)}, \label{eq:perano}
\eeq
with $\g_A=-2\g_Q$. It is also easy to check that the fixed point is infrared stable.
We see that the coupling constants (or more precisely, 't~Hooft couplings) of the theory are small in the above limit, and hence we can trust perturbative calculation.
Thus we consider that the existence of an infrared conformal fixed point in the above limit is established, and the fixed point values of the anomalous dimensions are
given by Eq.~(\ref{eq:perano}).

However, we are interested in the case where the coupling is strong, so that the anomalous dimension $\g_Q$ is quite large and any perturbative calculation
is not reliable at all.
A very astonishing fact of supersymmetric conformal field theory is that anomalous dimensions of fundamental
fields can be determined exactly, even in strongly coupled theories. The general method is called $a$-maximization~\cite{Intriligator:2003jj}.
In ${\cal N}=1$ superconformal field theories, there is an $R$ symmetry which appears in superconformal algebra (which is an extension of ordinary supersymmetry algebra). In some theories there may be a unique 
anomaly free $R$ symmetry, and if the theories are in the conformal window~\cite{Seiberg:1994pq}, 
the $R$ symmetry must be the one which appears in the superconformal algebra. However, in general there is a family of 
anomaly free $R$ symmetries (as in the model of Section \ref{sec:2}; see below), and we cannot determine  from symmetry argument alone which $R$ symmetry is the superconformal one~\footnote{In this paper we do not consider the case where the superconformal $R$ symmetry is an accidental symmetry of a low energy theory.}. 
According to Ref.~\cite{Intriligator:2003jj}, the superconformal $R$ symmetry is the one which locally maximizes the following combination of
t'~Hooft anomalies,
\beq
\sum_i [3(R_i-1)^3-(R_i-1)],
\eeq
where the sum is taken over fermions of a theory and $R_i-1$ is the $R$ charge of the fermion in chiral field $i$. This condition determines the $R$ charge
$R_i$ of the chiral field $i$. Furthermore, the scaling dimension $D_i$ and the anomalous dimension $\g_i$ of chiral field $i$ is related to the $R$ charge
$R_i$ by the equation
\beq
1+\frac{\g_i}{2}=D_i=\frac{3}{2}R_i. \label{eq:confrelation}
\eeq
The first equality in Eq.~(\ref{eq:confrelation}) is almost the definition of the anomalous dimension in conformal field theory. 
For the second equality, see e.g.~\cite{Minwalla:1997ka}. 
From Eq.~(\ref{eq:confrelation}), we can determine the anomalous dimension $\g_i$ from the $R$ charge $R_i$.

\begin{table}[Ht]
\begin{center}
\begin{tabular}{|c|c|c|c|}
\hline 
&$A$&$Q,~\tilde{Q}$&$P,~\tilde{P}$ \\ \hline
$R$&$-2x$&$1+x$&$1+N_P^{-1}(2N_C-N_Q)x$ \\ \hline
\end{tabular}
\caption{$R$ charges of the fields. $x$ is a parameter that parametrizes the ambiguity of the definition of $R$ symmetry.}
\label{table:app1} 
\end{center}
\end{table}

Let us apply the above method to the model of Section \ref{sec:2}. We, here, neglect all masses for the fields, and assume that 
the model is in the conformal window~\cite{Seiberg:1994pq} for certain values of $N_C,~N_Q$ and $N_P$.
The $R$ charges of the fields are shown in Table \ref{table:app1}. We have imposed that $Q$ and $\tilde{Q}$ ($P$ and $\tilde{P}$) have the same $R$ charge. Even then, the $R$ charges of the fields are not
uniquely determined. We have parametrized the ambiguity of $R$ charges by $x$. Then, we define the following function of $x$:
\beq
a(x)&\equiv& \sum_i [3(R_i-1)^3-(R_i-1)] \nonumber \\
&=&(N_C^2-1)[3(-2x-1)^3-(-2x-1)]+2N_CN_Q[3x^3-x]  \nonumber \\
&&+2N_CN_P[3(N_P^{-1}(2N_C-N_Q)x)^3-(N_P^{-1}(2N_C-N_Q)x)] .
\eeq
Then, the condition for the local maximization of $a(x)$ is given by
\beq
\frac{\q a(x)}{\q x}=0,~~~\frac{\q^2 a(x)}{\q x^2}<0.
\eeq
Solving these equations is quite straightforward. 
Using the solution for $x$, the anomalous dimension of, e.g.,~$Q$ is given by $\g_Q=3R_Q-2=3x+1$. 
The value of $\g_Q$ is listed in Table~\ref{table:1}.

\begin{table}[Ht]
\begin{center}
\begin{tabular}{|c|c|c|c|c|c|}
\hline 
&$N_P=2$&$N_P=3$&$N_P=4$&$N_P=5$&$N_P=6$ \\ \hline
$N_C=5$&0.273&0.143&0.059&$\times$&$\times$  \\ \hline
$N_C=6$&0.422&0.280&0.179&0.104&0.046 \\ \hline
$N_C=7$&$\times$&0.391&0.287&0.205&0.138 \\ \hline
$N_C=8$&$\times$&0.478&0.376&0.292&0.223 \\ \hline
$N_C=9$&$\times$&$\times$&0.448&0.366&0.296 \\ \hline
\end{tabular}
\caption{The values of $\g_Q$ obtained in Eq.~(\ref{eq:perano}) by perturbative calculation. This table should be compared with Table~\ref{table:1}, where 
the exact value of $\g_Q$ is listed. $N_Q$ is taken to be 5.}
\label{table:4} 
\end{center}
\end{table}

As a check, we list the one-loop value of $\g_Q$ obtained in Eq.~(\ref{eq:perano}) in Table~\ref{table:4}.
Note that the agreement between Tables~\ref{table:1} and \ref{table:4} is quite good, and becomes better as the coupling becomes weaker as expected. 

For what values of $(N_C,N_Q,N_P)$ the model has a conformal fixed point is a rather nontrivial question.
In conformal field theory, it is known that all gauge invariant operators of a theory have scaling dimensions greater than or equal to $1$~\cite{Mack:1975je}.
As a criterion of the existence of a conformal fixed point, we require that all gauge invariant chiral (primary) operators have scaling dimensions greater than 
or equal to $1$. Such a criterion was first used in Ref.~\cite{Seiberg:1994pq} to show the presence of a conformal window in SUSY QCD.
Especially, in the case of the present models, we have imposed that the scaling dimensions of gauge invariant chiral operators $\tr A^2$ and $P^a\tilde{P}_b$
satisfy the conditions
\beq
D_{\tr A^2}&=&2\left(1+\frac{\g_A}{2}\right) \geq 1, \label{eq:unitbound1} \\
D_{P\tilde{P}}&=&2\left(1+\frac{\g_P}{2}\right) \geq 1. \label{eq:unitbound2}
\eeq 
In this paper we have assumed that if Eqs.~(\ref{eq:unitbound1},\ref{eq:unitbound2}) are satisfied, and if the theory is asymptotic free at the one-loop level, 
the model has a conformal fixed point~\cite{Seiberg:1994pq,Ibe:2005pj}. 
We see that the models in Table~\ref{table:1} without $\times$ indeed satisfy those conditions.

\end{document}